# Multistable Curved-Crease Origami Blocks for Reconfigurable Modular Building System

Munkyun LEE[a], Joseph M. GATTAS[b], Tomohiro TACHI[*]

[*] Department of General System Studies, Graduate School of Arts and Sciences, The University of Tokyo
153-8902, Building 15-702, 3-8-1 Komaba, Meguro-ku, Tokyo, Japan
tachi@idea.c.u-tokyo.ac.jp

[a] Department of Architecture, Graduate School of Engineering, The University of Tokyo

[b] School of Civil Engineering, University of Queensland

## Abstract

This study proposes a reconfigurable modular building system that assembles multistable curved-crease origami blocks. Curved-crease origami is designed with even-vertex polygonal trajectories and an elastica curvature profile. We then connect the matching ends to impart multistability. Through this design approach, we create various blocks and investigate their snapping and load-bearing behavior using finite element analysis. We design block assemblies of multi-story and quasi-continuous wall surfaces and fabricate a series of desktop and large-scale prototypes to demonstrate the flexibility and adaptability of our system for architectural use. Furthermore, by introducing a tension cable to the assembly, the assembled modules can be snapped into multiple configurations.

## 1. Introduction

### 1.1. Curved-Crease Origami

Curved-crease origami consists of curved creases and bendable panels that generate smooth surface curvature during deployment. This combination of folding and bending has demonstrated outstanding structural properties, including enhanced bending capacity, buckling resistance, and stress stiffening. These advantages have naturally attracted researchers and engineers in the civil and architectural realms, leading to large-scale applications. Such large-scale applications are particularly prominent in cylindrical topologies, with proposed applications including column design [1, 2], deployable space reflectors [3], beams [4], bridges [5, 6], self-locking coiled tubes [7], and curved wall modules [8].

Since these curved-crease origami inherently store bending stress in their panels, they require additional fixing equipment to maintain a deployed shape. This demands a design trade-off between shape stability and deployability, which is an essential challenge for large-scale applications. To tackle this issue, this paper extends upon the wall modules previously proposed by Lee et al. [8] and aims to introduce the multistability and enhance the reconfigurability of modular curved-crease origami structural systems.

### 1.2. Multistable Origami

Multistable origami possesses energy stability across multiple configurations during deployment. Several existing multistable origami design methods demonstrate that stiffness and stable shapes can be designed based on the rigid origami mechanism. The geometric incompatibility method [9, 10] combines multiple rigid origami with incompatible deployment paths to induce multistability and is suited for the stiffness design. The loop-closing method [11, 12] overconstrains a one-degree-of-freedom rigid origami strip by hinge-connecting its matching ends and is suited for stable shape design.

Recently, multistable origami based on curved-crease origami has also emerged, with proposed techniques including creating incompatibility between a linkage and curved origami [13], mode bifurcation [14], and embedding linear creases in concave surfaces of closed cylindrical curved-crease origami [6]. These suggest the potential for achieving self-supporting curved-crease origami by multistability. However, the knowledge of systematic design strategies and detailed mechanical behavior analyses for multistable curved-crease origami remains limited.

### 1.3. Objective

We present the design and analysis of multistable curved-crease origami blocks, termed CC-Blocks, and introduce a modular building system capable of generating reconfigurable, quasi-continuous, and self-supporting wall surfaces. Specific study objectives are to: (1) establish the CC-Block design method via a loop-closing method applied to cylindrical curved-crease origami, (2) analyze mechanical behaviors across various geometric parameters, and (3) demonstrate practical feasibility through both desktop- and large-scale prototypes. These objectives are addressed in Sections 2, 3, and 4, respectively. Our modular system offers a novel paradigm for lightweight, rapid, and reconfigurable construction, with potential applications in shelters, adaptive spatial structures, and facades.

## 2. Geometry Design

In this section, we present the design of multistable curved-crease origami blocks. We generate the cylindrical curved-crease origami by mirror-reflecting the profile curve along the polygonal trajectory. To introduce multistability, we used a *loop-closing* method that connects both ends of the cylindrical form. The resulting configuration shows stability in both flat-folded and deployed states.

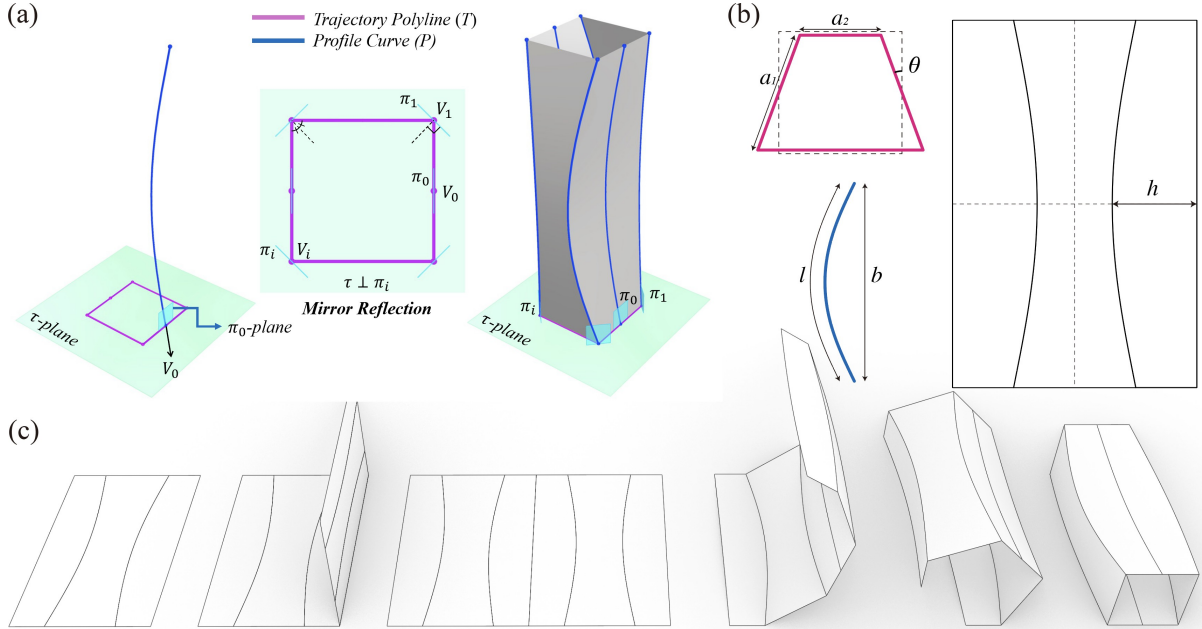

Figure 1: (a) Design process, (b) design parameters, and (c) deployment motion of curved-crease origami.

### 2.1. Curved-Crease Origami

Following the design method in [8, 15, 16], we construct cylindrical curved-crease origami using a *trajectory polyline* and a *profile curve*, as illustrated in Figure 1(a). Here, we define each vertex of the trajectory polyline as $V_i$ ($i \in [0, n]$). Reflection planes from $\pi_0$ to $\pi_i$ are generated at polyline vertices with normal vectors of planes coincident with the bisector of the corner angle (Figure 1(a)).

The profile curve is modeled as an elastica curve, which describes the naturally stable post-buckled shape of a slender elastic rod. Such elastica profiles have been shown to accurately represent the geometry of elastically bent curved-crease origami with an enforced, fixed-length boundary condition [16]. In our case, the profile is defined as a pinned–pinned elastica segment, characterized by an arc length l and a deformed height b, corresponding to the first mode configuration (Figure 1(b)). In this paper, we define the profile curvature (c) as the ratio $c = b/l$. To construct the curved-crease origami block, this profile is placed on the initial reflection plane $V_0$, and then mirrored successively across the adjacent planes along the trajectory, as illustrated in Figure 1(a)-right. The motion of the designed curved-crease origami is illustrated in Figure 1(c).

## 2.2. Loop-closing

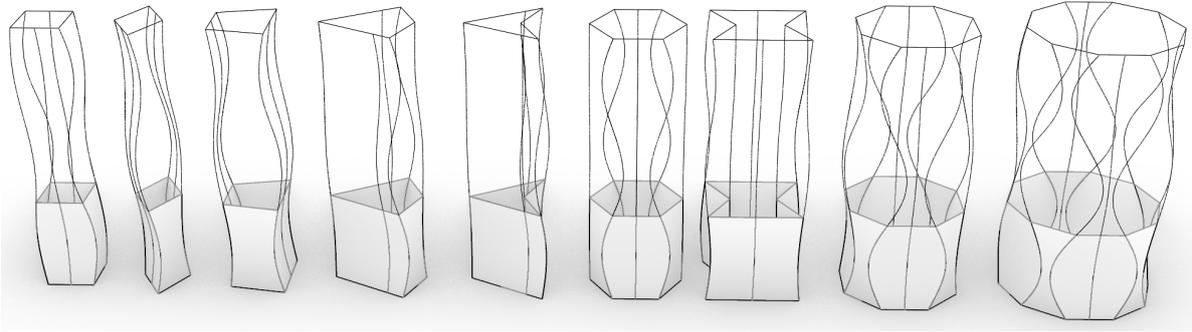

Figure 2: Design variations of CC-blocks with various trajectory polylines and k-mode elastica profile curves.

To achieve multistability in the designed cylindrical curved-crease origami, we adopt the loop-closing method [11, 12] (Section 1.2.). While the method has primarily been used for rigid origami, we extend this method to curved-crease origami. To get the closure cylindrical shape at the deployed and flat states: (1) The start vertex ($V_0$) and the end vertex ($V_n$) of the trajectory polyline should be matched to complete the loop. (2) The trajectory vertices should be an even number to allow connectivity and continuity of the curved surface. (3) At least two vertices should lie on straight edges rather than being corners, to create linear creases. And one of the edge-on vertices is set as $V_0 (= V_n)$. Finally, (4) this trajectory must be flat-foldable by folding these edge-on vertices in the outward direction. By following the design process (Figure 1(a)) while satisfying the above trajectory conditions, multistability can be achieved.

Figure 3(c) shows the deployment motion of the CC-block. While the motion deviates from the original curved-folding motion (Figure 1(c)), it still reaches both flat and deployed states. Compared to the original curved-folding, an increased panel distortion generates a distinct energy barrier between the two configurations.

Based on this design method, we can design flat-foldable and multistable CC-blocks with various trajectories and k-mode elastica profiles; such design variation is illustrated in Figure 2. Although various design configurations can be considered, in this paper, we limit our scope to first-mode elastica profiles to simplify the mechanical analysis and to trapezoidal trajectories to keep consistency with the assembly method employed in our previous work [8].

The column with the first-mode elastica profile has alternating convex and concave sides. In our design, we choose to put the linear creases on the concave side, similarly to [6]. This is because, through physical models, we observed no snapping when the creases are added to the convex side [3].

## 3. Mechanics

Here, we analyze the mechanical behaviors of CC-blocks across various design parameters and loading conditions through the finite element method, using *Abaqus 2024 (Dassault Systems)*. We consider a flat-folded tube as shown in Figure 3, where it lies on the *xy*-plane. When deploying, we have two options to apply compression to the *x*-direction (pinching) or tension to the *z*-direction (pulling). In Section 3.2., we analyze the snapping behaviors based on the pinching deployment and elaborate on the effect of different parameters on the snapping behavior. Additionally, we discuss the differences in snapping behavior when objects are pinched or pulled. In Section 3.3., we analyze the load-bearing behavior of the deployed module under compression in y directions, the typical governing load condition when used as a column or wall element.

## 3.1. Modeling and Boundary Conditions

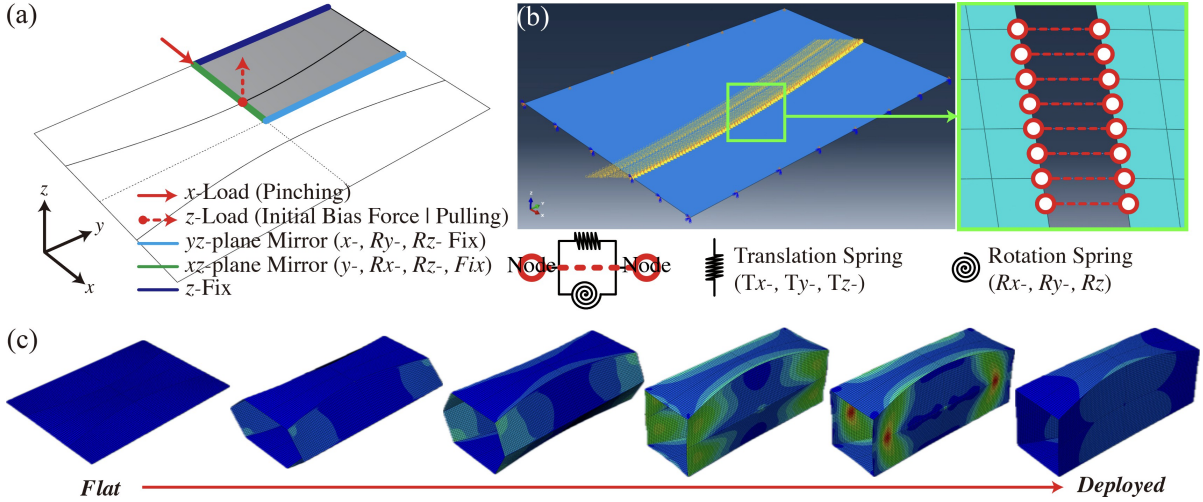

Figure 3: (a) Applied boundary conditions for snapping analysis, (b) Curved-crease modeling methods implemented in the FE software, (c) Deployment simulation and stress contour of the reference model of CC block.

Figure 3(a) shows the boundary conditions applied to the model. We used a mirror-symmetric condition to simplify the model design and reduce computational costs. The analysis of Figure 4(b, d, e, f, i), which considers triaxially-symmetric CC blocks with a rectangular trajectory, 1/8 of the full block was modeled. For the analysis of Figure 4(c, g, h), which considers a biaxially-symmetric CC block with a trapezoidal trajectory, 1/4 of the full block was modeled. The analysis models are initialized in the flat state and constrained with mirror symmetry. The vertical mid-section along the *yz*-plane is constrained with an *x*-translation fix and restrictions on $R_y$ and $R_z$ rotations. The horizontal mid-section along the *xz*-plane is constrained with a *y*-translation fix and restrictions on $R_x$ and $R_z$ rotations. The outermost edges along the *y*-axis, corresponding to the midpoint joints of the trajectory, are fixed only in the *z*-direction, as this allows ideal hinge behavior without introducing rotational stiffness.

Figure 3(b) shows the modeling strategy used to simulate the CC-block with finite element analysis (FEA). Each panel was modeled using S8R quadrilateral mesh shell elements. Adjacent edges with equal length were divided into the same number of nodes to ensure node connectivity. To simulate the curved folding behavior in the FEA software, we developed a new hinge modeling method, which is based on the pin-joint array hinge proposed in [17]. We implemented pin joints by connecting mesh nodes at identical locations using connector elements. Each connector type was defined with Cartesian-type for the translational spring in x, y, and z-direction and Rotation-type for the rotational spring in $R_x$, $R_y$, and $R_z$-direction (Figure 3(b)). By assigning equal stiffness to the *x*, *y*, and *z* translational directions, and likewise to the $R_x$, $R_y$, and $R_z$ rotational directions, we ensured that each pin-joint maintains the intended translational and rotational stiffness regardless of its position in the global coordinate system. This approach differs from previous methods [18, 19, 20], which rely on local coordinate systems at each node connection. In contrast, our method achieves hinge behavior using only a global coordinate, simplifying the modeling process and reducing computation cost.

In this paper, we focus only on the geometric and elastic behavior of the structure. Complex physical phenomena from panels and creases, such as plasticity, viscoelasticity, damping, and contact, are not considered. As such, we set linear stiffness values for each pin joint with a translational stiffness of "$1.0 \times 10^6 \; N/mm/edge\; nodes\; count$" and a relatively small rotational stiffness of "$0.001 \; N \cdot mm/rad/edge\; nodes\; count$".

For the snapping analysis, forced displacements of $h$ value were applied along the *x*-direction at the center point of the concave panel. For breaking the initial flat configuration, we applied a $1\;N$ initial bias load in the *z*-direction for $0 - 0.1\;s$, after which it was released. For the material setting, we used polypropylene material properties with an elastic modulus of $1134 \; MPa$, density of $9.05 \times 10^{-10} ton/mm^3$, and the Poisson's ratio of 0.38. The snapping analysis is performed using a quasi-static method in the dynamic implicit solver.

### 3.2. Snapping Analysis

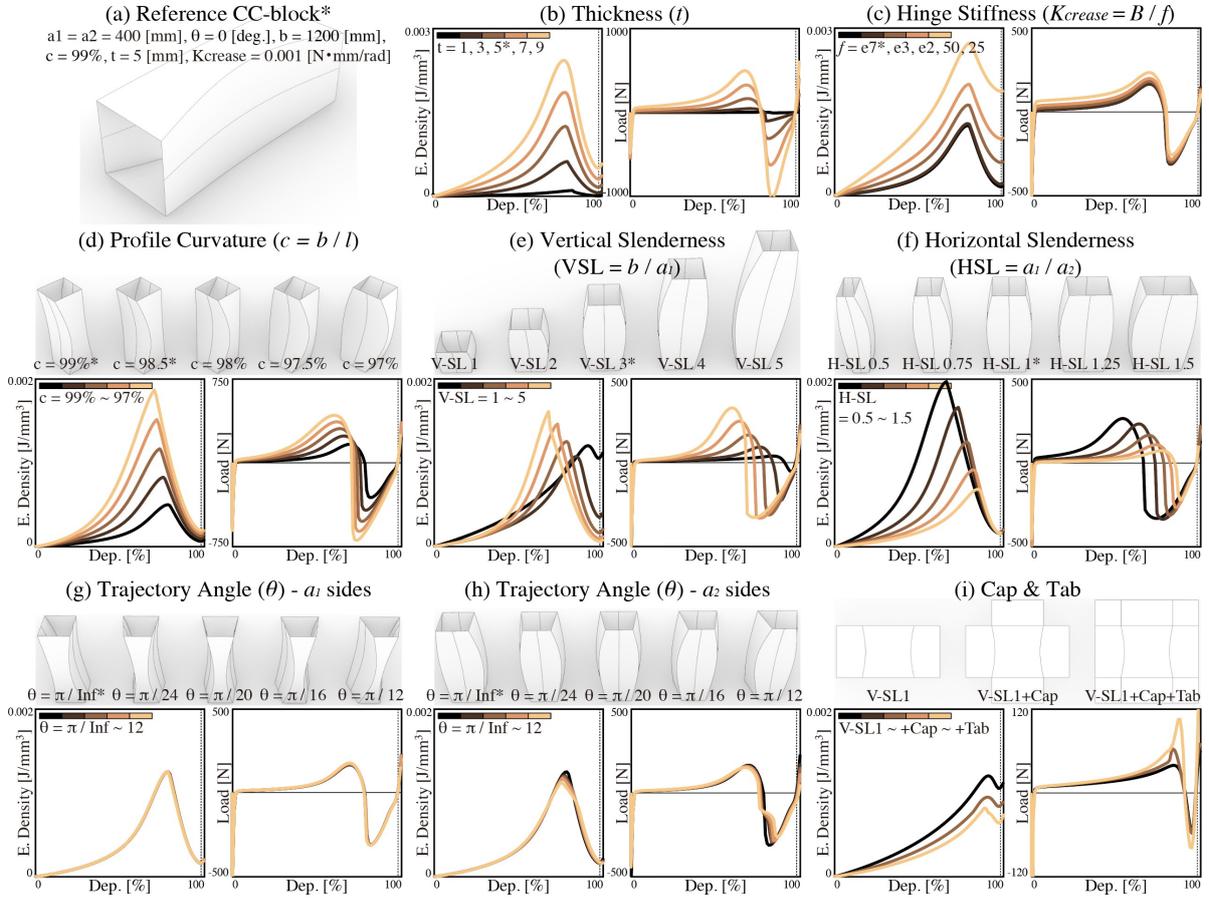

Figure 4: (a) Reference CC-block and design parameter values. Analysis results of (b) thickness parameters, (c) hinge stiffness parameters, (d) vertical slenderness parameters, (e) elastica profile curvature parameters, (f) horizontal slenderness parameters, (g) trajectory angle parameters which the additional creases located in the $a_1$-side panels, or (h) $a_2$-side panels. (i) Behavior differences with and without the cap and tab.

Figure 4(b–i) show the FEA results of the pinching loading behavior depending on design parameters. The horizontal axis (deployment) of plots is defined as the forced-displacement ($d$) divided by the insertion depth ($h$) of the concave panel in deployed states, $d/h$. The left plots show the energy density, dividing total strain energy by the volume of the geometry, and the right plots show the required force during deployment. By comparing the energy density, we can exclude the effect of panel size difference along the design parameters, providing a more uniform comparison across different CC-block

geometries. Note that energy density plots in Figure 4 have a small initial peak, which occurs due to the initial bias *z*-load, and because of this, the negative force is captured in the initial step of the force plots. We set the reference model for the comparison, which is shown with design parameter values in Figure 4(a). The asterisk ∗ marked in figures or legends represents this reference model. Further result details can be found in Table 1– 3. Table 1 presents the deployment locations of energy density peaks. Table 2 shows the energy ratio between the peak energy and the second stable state energy value ($E_{peak}/E_{stable}$), indicating how stable a model is at the deployed state. Table 3 shows the absolute ratio between maximum and minimum force values ($|F_{min}/F_{max}|$), representing the required load difference for deploying and folding the CC-block.

Table 1: Peak locations.

| Peak Dep. [%] | Fig.3(b) | Fig.3(c) | Fig.3(d) | Fig.3(e) | Fig.3(f) | Fig.3(g) | Fig.3(h) | Fig.3(i) |
|---|---|---|---|---|---|---|---|---|
| $C_{Black}$ | 84.204 | 79.947 | 80.046 | 92.665 | 67.540 | 80.046 | 80.05 | 92.665 |
| $C_{DarkBrown}$ | 81.085 | 79.947 | 77.162 | 85.079 | 74.737 | 79.813 | 78.61 | – |
| $C_{Brown}$ | 80.046 | 79.947 | 75.035 | 80.046 | 80.046 | 79.293 | 78.66 | 92.665 |
| $C_{LightBrown}$ | 79.006 | 80.972 | 73.291 | 74.514 | 83.607 | 79.859 | 77.6 | – |
| $C_{Yellow}$ | 79.006 | 80.972 | 71.735 | 69.062 | 86.559 | 80.024 | 76.66 | 90.606 |

Table 2: Energy ratio ($E_{peak}/E_{stable}$).

| $E_{peak}/E_{stable}$ | Fig.3(b) | Fig.3(c) | Fig.3(d) | Fig.3(e) | Fig.3(f) | Fig.3(g) | Fig.3(h) | Fig.3(i) |
|---|---|---|---|---|---|---|---|---|
| $C_{Black}$ | 14.934 | 7.739 | 7.821 | 1.142 | 12.344 | 7.821 | 7.821 | 1.142 |
| $C_{DarkBrown}$ | 10.414 | 6.289 | 8.586 | 3.073 | 10.309 | 7.748 | 7.515 | – |
| $C_{Brown}$ | 7.821 | 2.738 | 9.063 | 7.821 | 7.821 | 7.745 | 7.427 | 1.133 |
| $C_{LightBrown}$ | 6.051 | 1.950 | 9.421 | 15.904 | 5.736 | 7.735 | 7.345 | – |
| $C_{Yellow}$ | 4.853 | 1.463 | 9.620 | 27.139 | 4.321 | 7.649 | 6.879 | 1.284 |

Table 3: Force ratio between deployment and folding ($|F_{min}/F_{max}|$).

| $|F_{min}/F_{max}|$ | Fig.3(b) | Fig.3(c) | Fig.3(d) | Fig.3(e) | Fig.3(f) | Fig.3(g) | Fig.3(h) | Fig.3(i) |
|---|---|---|---|---|---|---|---|---|
| $C_{Black}$ | 2.723 | 1.886 | 1.886 | 1.240 | 1.245 | 1.886 | 1.886 | 1.240 |
| $C_{DarkBrown}$ | 2.252 | 1.858 | 1.777 | 2.442 | 1.441 | 1.863 | 1.843 | – |
| $C_{Brown}$ | 1.886 | 1.635 | 1.719 | 1.886 | 1.886 | 1.853 | 1.830 | 0.862 |
| $C_{LightBrown}$ | 1.912 | 1.428 | 1.675 | 1.338 | 2.440 | 1.832 | 1.802 | – |
| $C_{Yellow}$ | 2.022 | 1.093 | 1.633 | 1.001 | 2.981 | 1.793 | 1.753 | 0.738 |

Figure 4(b) shows the influence of panel thickness (*t*). As thickness increases, the peak appears slightly earlier (Table 1) and becomes more rounded, which can weaken the snapping behavior. Although both the energy barrier and stable state energy rise, the stability in the deployed state is reduced (Table 2). This is due to the thicker panel requiring more energy at the same curvature. Therefore, the thicker panels demand greater force during deployment and folding. While folding generally demands greater force than deployment, no consistent trend is observed with varying thickness (Table 3).

Figure 4(c) shows the influence of hinge stiffness. The hinge stiffness parameter is defined as the bending rigidity ($B = Et^3/12 /(1 - v^2)$) of panels divided by a length-scale factor *f*. As stiffness increases, although both the energy barrier and stable state energy rise, the stability in the deployed state is reduced (Table 2). Energy peaks are almost unaffected by hinge stiffness (Table 1). As the hinge stiffness increases, the difference in the required force between deployment and folding decreases (Table 3). Therefore, excessive stiffness may eliminate multistability, leading to a monotonic increase in energy.

Figure 4(d) shows the influence of profile curvature ($c = b/l$) with a fixed height $b = 1200$. A higher *c* raises the energy barrier and shifts the load peak earlier (Table 1), while the stable state energy remains similar, indicating improved stability (Table 2). While the larger curvature increases the required forces, the difference in the required force between deployment and folding decreases (Table 3).

Figure 4(e) shows the influence of the height-to-edge length ratio ($b/a_1$) while maintaining a constant $c$ of 99%. When the vertical slenderness (V-SL) increases, there is a higher energy barrier with a sharper and earlier peak during deployment (Table 1). The energy density at the stable state decreases, enhancing stability in the deployed state (Table 2). Absolute force values become greater, while the force difference between deployment and folding becomes smaller (Table 3).

Figure 4(f) shows the influence of the trajectory edge length ratio ($a_1/a_2$) while maintaining a $c$ of 99%. A smaller horizontal slenderness (H-SL) leads to a higher energy barrier with earlier (Table 1) and sharper peaks, indicating clear snapping behavior. Since the energy density at the stable state is the same, the stability in the deployed state is improved (Table 2). Additionally, while the required force increases for both deployment and folding, the force difference between the two becomes smaller (Table 3).

Figures 4(g, h) show the influence of trajectory angle ($\theta$). In (h), where the linear crease is on the $a_1$ side, $\theta$ has almost no effect. In (g), with the crease on the $a_2$ side, a geometric difference between front and back panels is observed. So, plots show the averaged energy density and load. As $\theta$ increases, the energy peak very slightly decreases. In general, a larger $\theta$ slightly reduces stability (Table 2) at the deployed state and the force ratio between deployment and folding (Table 3). The slight kink in the force curve reflects differences in snap-through timing and energy barriers between the front and back panels, indicating multistability in this geometry.

Figure 4(i) compares the behavior of the least stable case (V-SL1) with and without cap and tab additions. Adding only a cap reduces the stability at the deployed state and increases the required deployment force, but the folding force remains similar. Adding both the cap and tab improves stability (Table 2) and increases the required forces for both deployment and folding, while still maintaining a lower folding force compared to deployment (Table 3). Thus, using both cap and tab is recommended to ensure greater stability in the deployed state.

Figure 5(a) compares the snapping behavior of the reference CC-block under *x*-direction (pinching) and *z*-direction (pulling) loading, showing energy density (left) and force (right) plots. The *x*-load boundary condition is shown in Figure 3(a), while the *z*-load applies the forced-displacement only along the *z*-axis. The same geometry results in identical energy at both the peak and deployed states. However, *z*-loading produces a delayed and sharper energy peak, resulting in more distinct snapping behavior. It also requires higher deployment and folding forces, with a force ratio ($|F_{min}/F_{max}|$) of 3.769, nearly double that of the *x*-load (1.886). These results suggest that the CC-block in the *z*-direction has better resistance to flattening, emphasizing the need to consider loading direction in CC-block assembly and applications.

### 3.3. Load-bearing Analysis

Figure 5(b) and (c) illustrate the load-bearing behavior of the deployed reference model under *y*-direction compression. The simulation begins with deployment from the flat state, followed by boundary release to reach a naturally stable configuration. Then, the top and bottom edges are fixed, and a 250 mm compressive forced-displacement is applied. Here, we added a material property of 1% Rayleigh (both mass and stiffness) damping based on the first eigenmode for the stable analysis. For comparison, a simple fixed square column was also analyzed under the same conditions. The compression analysis is performed using a quasi-static method in the dynamic implicit solver, and the compression motion is shown in Figure 5(c).

Figure 5(b) presents the energy density (left) and force (right) plots. Compared to the square column, the CC-block shows earlier buckling and lower load capacity. However, due to its curved panel, the snapping behavior seems significantly mitigated, and the post-buckling load remains nearly constant, with an elastic buckling failure mode and load similar to that of the square column. Buckling occurs at the mid-height of the CC-block, where curvature (stress) from panel bending is highest (Figure 5(c)).

Up to a displacement of 100 mm, panels do not intersect, but beyond this point, back panel intrusion occurs, indicating that panel contact could further influence actual behavior.

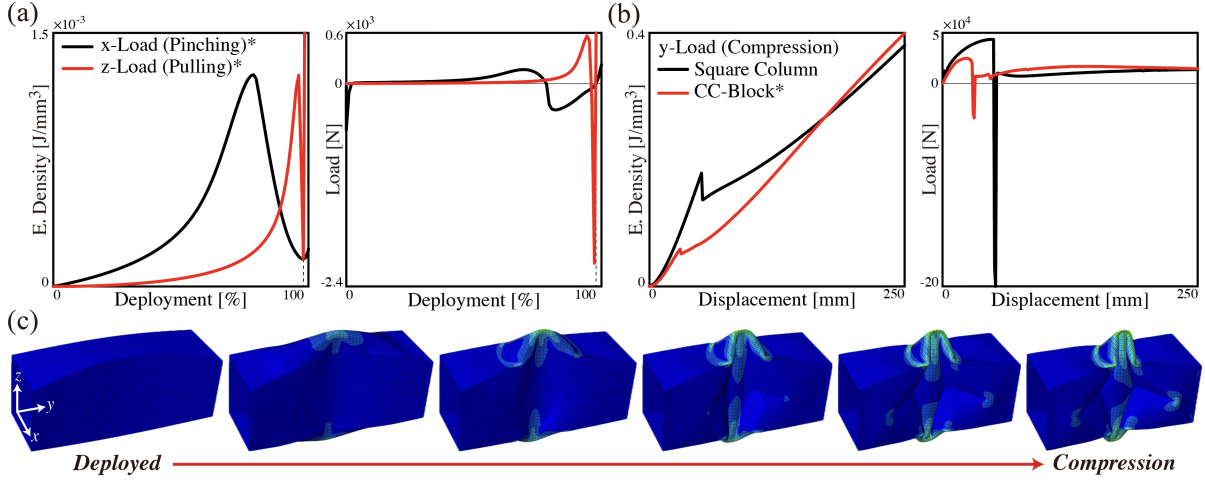

Figure 5: (a) Snapping behavior under different loading directions. (b) Compressive load-bearing behavior under *y*-direction compression and comparison with a square column. (c) Compression simulation and stress contour.

## 4. Fabrication and Application

Here, we fabricate both desktop- and large-scale CC-block prototypes to validate their multistability, modular assembly, and feasibility for application as a modular building system.

### 4.1. Prototype: CC-Column

#### *4.1.1. Desktop-scale Implementation*

The CC-column design is adapted from the desktop-scale prototype presented in [8]. While maintaining the original geometry, we apply the loop-closing method to transform the block into a multistable, flat-foldable configuration. The design parameters (Figure 1(b)) for trapezoidal trajectory parameters as $a_1 = a_2 = 46mm$, with an angle $\theta = 11.25\ deg$.; profile parameters for a first-mode elastica with $b = 118.8\ mm$ and $l = 120\ mm$, for $c = 99\%$. Here, we utilize the second-mode elastica by doubling $b$ and $l$, while keeping $c$.

For the desktop-scale fabrication, we used a 0.5 mm thick polypropylene-coated paper ($\alpha Yupo$), and we attached non-woven fabric to the whole side face using the double-sided tape sheet. The fabric membranes work as hinges and have relatively low hinge stiffness. End caps and tabs are included. The fabrication process is as follows (see Figure 6(a)): (1) the pattern is cut into the $\alpha Yupo$ sheet using the laser cutting machine; and (2) the double-sided tape sheet is attached to one whole side along with the non-woven fabric membrane.

Figure 6(b) shows the deployment and folding motion of the CC-column. The deployment was achieved by pressing the concave region (like Figure 3(a)), while folding was performed by initiating compression from the convex region. Snap-through behavior was observed only in the concave regions. As analyzed in Figure 4(h), the prototype exhibits multistability through different snapping between the front and back panels. Qualitatively, model deployment was consistent with the results of (Figure 4(f)), with narrower concave panel side demanding greater deployment force and being more stable in the deployed state.

Figure 6(c, d) shows shape assembly rules and various assembled configurations using a single type of CC-column. The assembly logic follows the shape grammar approach previously introduced in [8]. In addition to the original rules (SR-1 to SR-3), we introduce a new vertical assembly rule, SR-0, which enables multi-story assembly. This addition extends the design range of the modular system by allowing vertical stacking with continuous wall surfaces, using the same geometry.

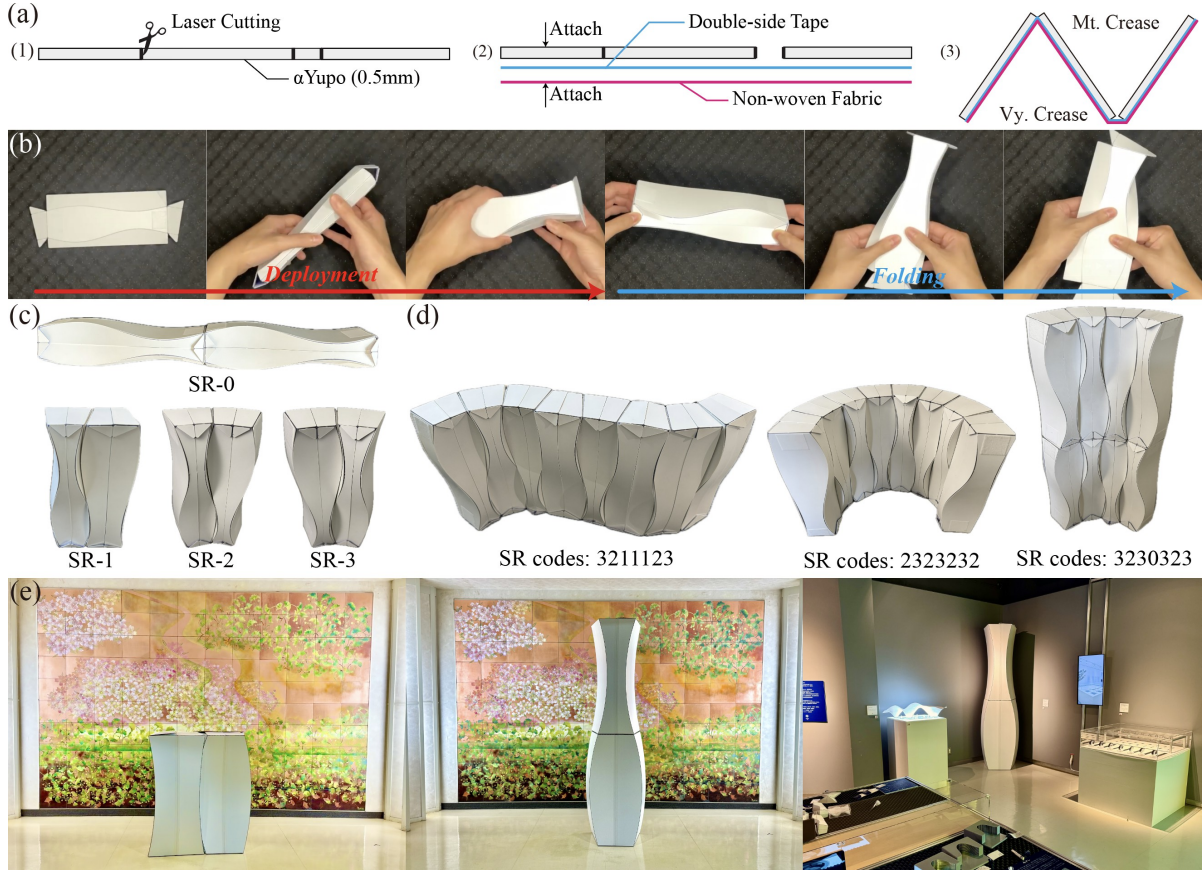

Figure 6: (a) Process of the desktop-scale fabrication. (b) Deployment and folding process of the desktop-scale CC-column. (c) Assembly rules and (d) Assembled configurations. (e) Large-scale CC- column.

*4.1.2. Large-scale Implementation*

Figure 6(e) presents the large-scale CC-column, assembled in both horizontal and vertical configurations, where the geometry parameters related to the length are scaled up by 10 from the desktop-scale. For the panel material, we employed a 3 mm-thick polypropylene bubble-core sandwich sheet *(Plapearl: PCPPZ-050, Kawakami sangyo Co., Itd.)*, selected for its lightweight and elastic properties. Due to panel material size limitations, the CC-column was fabricated by dividing it into quarters. We used the CNC cutter, and all curved creases were made with half-cuts, while linear creases were made with full-cuts and then reconnected using duct tape (*Gorilla Tape*). Each block has caps and tabs to fix the deployed state and to enable vertical stacking. The connection surfaces between modules were attached with Velcro tape. As observed in Section 2.2., only the block where the linear crease is located on the concave panel showed snapping behavior, while the other did not. To get snapping as well, the linear crease should be repositioned to the concave side (Figure 4(g)). Each block is rigid enough to bear the weight of another block in the deployed state. Both desktop and large-scale CC-column were exhibited at *Connecting Artifacts 04* in Tokyo, Japan (Figure 6(e)-right).

## 4.2. Prototype: Reconfigurable CC-Box

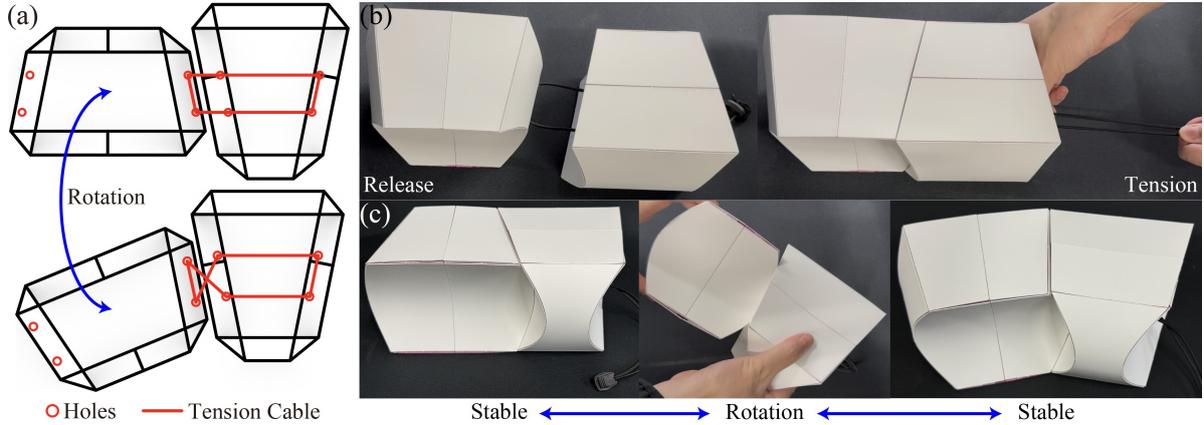

Figure 7: (a) Assembly method with tension cable, (b) CC-box prototypes and tension introducing, (c) motion of CC-box reconfiguring.

The design parameters (Figure 1(b)) for trapezoidal trajectory parameters as $a_1 = 100\ mm$, $a_2 = 80.491\ mm$, with an angle $\theta = 11.25\ deg$.; and profile curve parameters for a first-mode elastica with $b = 100\ mm$ and $l = 108.108\ mm$, for $c = 92.5\ \%$. End caps and tabs were included. For the reconfigurable prototype, we utilized two blocks as in Section 4.1.2., with linear creases placed on the concave panels. Here, we introduced a tension cable for secure block assembly while also allowing for relative rotation of adjacent blocks in the assembly. Figure 7(a) illustrates the concept, and it can be extended by inserting additional blocks in between. Figure 7(b) shows the process of introducing tension using rubber bands and stoppers. After introducing the tension, the blocks interlock rigidly. Figures 7(a, c) demonstrate reconfiguration between modules, showing how rotation enables a transition from linear (left) to curved (right) wall assemblies. An interesting observation is that the mismatch during the rotation causes cable extension, creating snapping behavior when the surfaces realign. This highlights that our new assembly method can achieve reconfigurability with multistability through simple rotation, without disassembly.

## 5. Conclusion and Future Works

In this study, we established a design methodology for achieving multistability in curved-crease origami and analyzed the snapping behavior under various geometric design parameters. Additionally, we investigated the influence of loading directions on snapping behavior and load-bearing capacity. Both desktop- and large-scale CC-block prototypes were fabricated, demonstrating the feasibility of our modular system, which is capable of forming multistory, quasi-continuous, curved wall surfaces. Furthermore, we enhanced the assembly method by introducing a tension cable, enabling post-deployment reconfigurability with snapping. Future work includes load-bearing evaluations and application-driven designs targeting actual usage cases (for example, emergency shelter).

## Acknowledgements

The first author was supported by JSPS 24KJ0648. This work is supported by JSPS KAKENHI 24H00822, JST AdCORP JPMJKB2302, and JST ASPIRE JPMJAP2401. We thank Kawakami Sangyo for providing materials and manufacturing of the large-scale model. We thank NIKKEN Sekkei for helpful discussions regarding the applications of modular systems. The second author would like to thank Qingyun Zhang for assistance in polypropylene material property evaluation.